\documentclass[times,12pt, twocolumn]{aastex631}

\usepackage{graphicx}		
\usepackage{latexsym, float}		
\usepackage{bm}			
\usepackage[english]{babel}
\usepackage{lipsum}
\usepackage{natbib}
\usepackage{appendix}
\usepackage{color}		
\newcommand{\lsim}{\mbox{\rlap{\hbox{\lower2pt\hbox{\ensuremath{\sim}}}}\raise2pt\hbox{\ensuremath{<}}}}%
\newcommand{\gsim}{\mbox{\rlap{\hbox{\lower2pt\hbox{\ensuremath{\sim}}}}\raise2pt\hbox{\ensuremath{>}}}}%
\renewcommand{\gtrsim}	{\ensuremath{\gsim}}
\renewcommand{\lesssim}	{\ensuremath{\lsim}}
\newcommand\eg           {{\it e.g.}, }

\newcommand\arcspt       {{$\buildrel{\prime\prime}\over .$}}
{\relax}%
\newcommand\n            {\noindent}

\newcommand\sn           {\smallskip\noindent}

\newcommand\degree       {{\ifmmode^\circ\else$^\circ$\fi}} 
\newcommand\arcm         {{\ifmmode {'\ }\else$'     $\fi}} 
\newcommand\arcs         {{\ifmmode{''\ }\else$''    $\fi}} 
\newcommand\cge          {{$\gtrsim$}}
\newcommand\cle          {{$\lesssim$}}

\newcommand\ie           {{\it i.e.},}

\newcommand\AV           {{$A_{\rm V}$}}

\newcommand\Ho           {{$H_{0}$} }

\newcommand\kmsMpc       {{km\ s$^{-1}$\ Mpc$^{-1}$} }

\newcommand\mum          {{\micron}}

\newcommand\Mo           {{M$_{\odot}$}}

\newcommand{\DELETED}[1]{\relax}%
{\relax}%

\usepackage{lineno}
\begin{document}
\title{JWST NIRCam Photometry: A Study of Globular Clusters Surrounding Bright Elliptical Galaxy VV~191a at z = 0.0513}

\author[0000-0001-6265-0541]{Jessica M. Berkheimer}
\affiliation{School of Earth \& Space Exploration, Arizona State University, Tempe, AZ\,85287-1404, USA}

\author[0000-0001-6650-2853]{Timothy Carleton}
\affiliation{School of Earth \& Space Exploration, Arizona State University, Tempe, AZ\,85287-1404, USA}

\author[0000-0001-8156-6281]{Rogier A.~Windhorst} 
\affiliation{School of Earth \& Space Exploration, Arizona State University, Tempe, AZ\,85287-1404, USA}
\affiliation{Department of Physics, Arizona State University, Tempe, AZ\,85287-1504, USA}

\author[0000-0002-6131-9539]{William C. Keel} 
\affiliation{Department of Physics and Astronomy, University of Alabama, Box 870324, Tuscaloosa, AL\,35404, USA}

\author[0000-0002-4884-6756]{Benne W. Holwerda} 
\affiliation{Department of Physics and Astronomy, University of Louisville, Louisville KY 40292, USA} 

\author[0000-0001-6342-9662]{Mario Nonino} 
\affiliation{INAF-Trieste Astronomical Observatory, Via Bazzoni 2, 34124, Trieste, Italy}

\author[0000-0003-3329-1337]{Seth H.~Cohen} 
\affiliation{School of Earth \& Space Exploration, Arizona State University, Tempe, AZ\,85287-1404, USA}

\author[0000-0003-1268-5230]{Rolf A.~Jansen} 
\affiliation{School of Earth \& Space Exploration, Arizona State University, Tempe, AZ\,85287-1404, USA}

\author[0000-0001-7410-7669]{Dan Coe} 
\affiliation{AURA for the European Space Agency (ESA), Space Telescope Science Institute, 3700 San Martin Drive, Baltimore,vvv MD\,21218, USA}

\author[0000-0003-1949-7638]{Christopher J. Conselice}
\affiliation{Jodrell Bank Centre for Astrophysics, Alan Turing Building, University of Manchester, Oxford Road, Manchester M13 9PL, UK}

\author[0000-0001-9491-7327]{Simon P. Driver} 
\affiliation{International Centre for Radio Astronomy Research (ICRAR) and the International Space Centre (ISC), The University of Western
Australia, M468, 35 Stirling Highway, Crawley, WA 6009, Australia}

\author[0000-0003-1625-8009]{Brenda L.~Frye} 
\affiliation{Department of Astronomy\,/\,Steward Observatory, University of Arizona, 933 N.\ Cherry Ave., Tucson, AZ\,85721, USA}

\author[0000-0001-9440-8872]{Norman A.~Grogin} 
\affiliation{Space Telescope Science Institute, 3700 San Martin Drive, Baltimore, MD\,21218, USA}

\author[0000-0002-6610-2048]{Anton M.~Koekemoer} 
\affiliation{Space Telescope Science Institute, 3700 San Martin Drive, Baltimore, MD\,21218, USA}

\author[0000-0003-1581-7825]{Ray A. Lucas} 
\affiliation{Space Telescope Science Institute, 3700 San Martin Drive, Baltimore, MD\,21218, USA}

\author[0000-0001-6434-7845]{Madeline A.~Marshall} 
\affiliation{National Research Council of Canada, Herzberg Astronomy \& Astrophysics Research Centre, 5071 West Saanich Road, Victoria, BC V9E\,2E7, Canada}

\author[0000-0003-3382-5941]{Nor Pirzkal} 
\affiliation{Space Telescope Science Institute, 3700 San Martin Drive, Baltimore, MD\,21218, USA}

\author[00000-0002-5404-1372]{Clayton Robertson} 
\affiliation{Department of Physics and Astronomy, University of Louisville, Louisville KY 40292, USA}

\author[0000-0003-0429-3579]{Aaron Robotham}
\affiliation{
International Centre for Radio Astronomy Research (ICRAR) and the International Space Centre (ISC), The University of Western
Australia, M468, 35 Stirling Highway, Crawley, WA 6009, Australia}

\author[0000-0003-0894-1588]{Russell E.~Ryan, Jr.} 
\affiliation{Space Telescope Science Institute, 3700 San Martin Drive, Baltimore, MD\,21218, USA}

\author[0000-0002-0648-1699]{Brent M. Smith}
\affiliation{School of Earth \& Space Exploration, Arizona State University, Tempe, AZ 85287-1404, USA}

\author[0000-0002-7265-7920]{Jake Summers}
\affiliation{School of Earth \& Space Exploration, Arizona State University, Tempe, AZ\,85287-1404, USA}

\author[0000-0001-9052-9837]{Scott Tompkins}
\affiliation{School of Earth \& Space Exploration, Arizona State University, Tempe, AZ\,85287-1404, USA}

\author[0000-0001-9262-9997]{Christopher N.A.~Willmer} 
\affiliation{Department of Astronomy\,/\,Steward Observatory, University of Arizona, 933 N.\ Cherry Ave., Tucson, AZ\,85721, USA}

\author[0000-0001-7592-7714]{Haojing Yan}
\affiliation{Department of Physics and Astronomy, University of Missouri, Columbia, MO\, 65211, USA}

\correspondingauthor{Jessica Berkheimer}
\email{jberkhei@asu.edu}

\shortauthors{Berkheimer et. al}
\shorttitle{Globular Clusters in VV~191a from JWST}

\begin{abstract}
James Webb Space Telescope NIRCam images have revealed 154 reliable globular cluster (GC) candidates around the $z = 0.0513$ elliptical galaxy VV~191a 
after subtracting 34 likely interlopers from background galaxies inside our search area. NIRCam broadband observations are made at 0.9--4.5 $\mu$m using the F090W, F150W, F356W, and F444W filters. Using PSF-matched photometry, the data are analyzed to present color-magnitude diagrams (CMDs) and color distributions that suggest a relatively uniform population of GCs, except for small fractions of reddest (5--8\%) and bluest (2--4\%) outliers. GC models in the F090W vs. (F090W--F150W) diagram fit the NIRCam data well and show that the majority of GCs detected have a mass of approximately $\sim$$10^{6.5}$$M_{\odot}$, with metallicities [Fe/H] spanning the typical range expected for GCs (--2.5\cle [Fe/H]\cle 0.5). However, the models predict $\sim$0.3--0.4 mag bluer (F356W--F444W) colors than the NIRCam data for a reasonable range of GC ages, metallicities, and reddening. Although our data does not quite reach the luminosity function turnover, the measured luminosity function is consistent with previous measurements, suggesting an estimated peak at $m_{\rm AB}$$\sim$--9.4 mag $\pm$ 0.2 mag in the F090W filter.

\end{abstract} \hspace{12pt}

\keywords{Instruments: James Webb Space Telescope--Techniques: photometry--Galaxies: globular clusters}

\section{Introduction}

\begin{figure*}
    \centering
    \includegraphics[width=\textwidth]{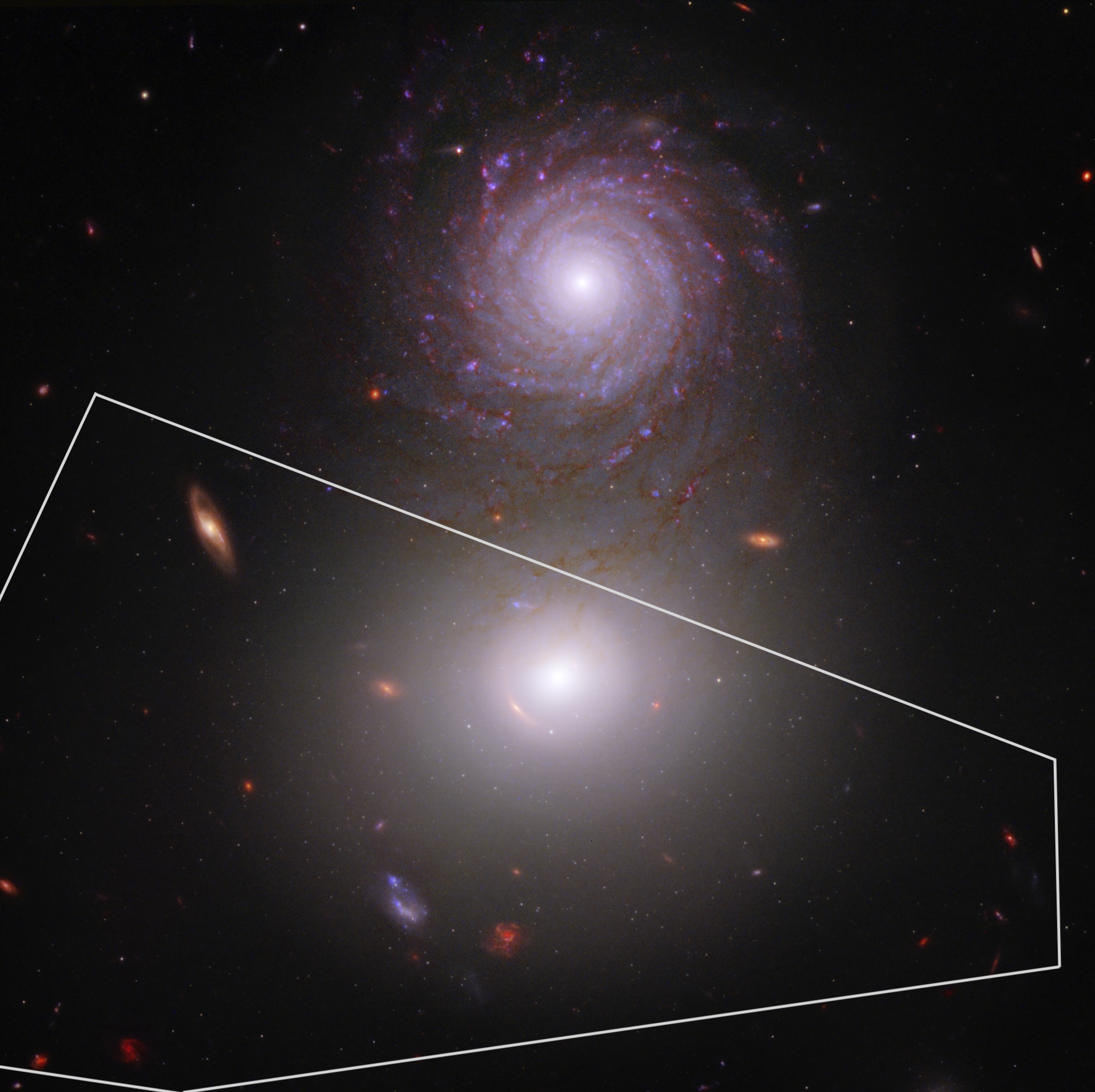}
    \caption{Image of galaxy pair VV~191a (elliptical galaxy on the bottom) and VV~191b (spiral galaxy on the top), which includes near-infrared light from Webb and ultraviolet and visible light from Hubble. In this image, green, yellow, and red were assigned to Webb’s near-infrared data in 0.9, 1.5, and 3.56 $\mu$m (F090W, F150W, and F356W, respectively). Blue was assigned to two Hubble filters, ultraviolet data taken in 0.34 $\mu$m (F336W), and visible light in 0.61 microns (F606W). Observations show an apparent population of GCs surrounding the elliptical {\citep{keel2023jwst}}. We added the white polygon to their image to outline the region useable for globular cluster searches in VV~191a, as determined in Fig.~2. 
}  
\label{fig:vv191}
\end{figure*}


Globular clusters (GCs) are among a galaxy's earliest and most pristine objects. They are stable, tightly bound collections typically ranging from $10^{4}$ to $10^{6}$ stars \citep[\eg][]{gratton2004abundance} and are intrinsically bright objects that can be observed at large distances. When considered as a system, they are excellent fossil records of the evolution and star-forming stages of the galaxy they surround \citep[\eg][]{ashman2008globular}. This view has been confirmed by direct measurements of their ages \citep[\eg][]{vandenberg2013ages}. Nearly all galaxies contain GCs, excluding dwarf galaxies below $10^7 M_\odot$ \citep[\eg][]{harris1991globular}. 

The number of GCs associated with a galaxy is observed to correlate with the galaxy's stellar mass, which itself correlates with dark matter halo total mass \citep[\eg][]{harris2017galactic,spitler2009new,georgiev2010globular,hudson2014dark}. Because ellipticals dominate the galaxy mass function, GCs are particularly abundant in these galaxies --- 1000s per galaxy in massive ellipticals \citep{harris1991globular}. Because elliptical galaxies have a much larger cluster density than spiral galaxies, they yield much better statistics \citep[\eg][]{kundu2001new}. With very little gas and dust obscuration and no star formation (SF) activity \citep[\eg][]{de2006formation}, it is relatively straightforward to distinguish GCs against the background of the elliptical host \citep{kundu2001new}.    

Broadly speaking, the current literature contains two categories of models for forming and evolving GC systems in the context of galaxy formation. The first category of models associates GC formation with special conditions in low-mass dark matter halos, during or before reionization \citep[\eg][]{peebles1968origin, katz2014clues}. The formation times for many GCs are thought to occur in the redshift range of $z \simeq 2$--8, and thus precede the bulk of SF in the universe \citep[\eg][]{hudson2014dark}. An early study by \cite{peebles1968origin} found that the Jeans mass at high redshift ($\sim$$10^{6} M_\odot$) is comparable to the GC stellar mass. More recent studies of GC formation at high redshift have proposed cluster formation within dark-matter halos of mass ${M}_{{{h}}}\lesssim {10}^{8}\;{M}_{\odot }$ with cooling driven by H$_2$ \citep[\eg][]{padoan1997star}. Because many GCs started forming before most stars in the universe, their formation may correlate more straightforwardly to the dark matter distribution than the bulk of the stellar mass in galaxies \citep{hudson2014dark}. \cite{blakeslee1997globular} suggested that the total numbers of GCs in bright cluster galaxies were directly proportional to the total dark-matter-dominated mass of the galaxy cluster.

The second category of models considers GC formation a natural byproduct of the active SF process and places high importance on galaxy mergers in producing GCs \citep[\eg][]{ ashman1992formation, zepf93}. This scenario focuses on somewhat lower redshifts. Present observations suggest that many clusters are formed during episodes of SF associated with a significant event during the formation process of the galaxy \citep{zepf1999globular, kundu2001new, linden2022goals}. This hypothesis can explain the increase in globular clusters around ellipticals \citep{ashman1992formation}. If elliptical galaxies are the aftermath of spiral galaxy mergers, GCs may form in such mergers \citep{schweizer1987star, ashman1992formation, whitmore1993hubble, zepf1999globular}. Theoretically, massive clusters form in merger events because of shocks and high turbulent pressure within interacting galaxies, which supports the formation of tightly bound clusters \citep[\eg][]{elmegreen1997universal}, and allows the formation of very massive gas clouds that are necessary to form GCs \citep{ashman1992formation}. 

One of the most interesting discoveries from observations of elliptical galaxies is the discovery of bimodal color distribution in many GCs \citep{zepf93, forbes1996hubble, kundu2001new}. Bimodal color and metallicity distributions are frequently interpreted as indicating two distinct modes of cluster formation. GCs found in the halo of progenitor spirals, like those observed in the Milky Way and M31, are expected to have a population of predominantly metal-poor clusters \citep{zepf1999globular}. The model predicted that the GC systems of elliptical galaxies would be composed of metal-poor populations from the progenitor spirals and metal-rich populations formed during the merger(s) that created the elliptical \citep{ashman1992formation, cote1998formation}. The metal-rich population will typically be significantly redder than the metal-poor population because metallicity differences dominate the broad-band colors of old GCs \citep{zepf93, ashman1998globular}. 

The older, metal-poor stellar population shows significant evidence suggesting a substantial age–metallicity degeneracy in GCs if only optical photometry is used \citep[\eg][]{worthey1994comprehensive, fan2006new}. The older stellar population is indistinguishable from that of a younger but more metal-rich population and vice versa \citep[][]{macarthur2004structure}. The optical colors of old populations are affected by the age-metallicity degeneracy, implying that the spectrophotometric properties of an unresolved stellar population cannot be distinguished from those of another population three times older and with half the metal content \citep{worthey1999age}. Adding infrared or near-infrared photometry can help break the age–metallicity degeneracy significantly \citep{wu2005optical, james2006optical}. 

This paper analyzes the globular cluster population in the nearby elliptical VV~191a. Visual inspection of the JWST infrared data presented first in \citet{keel2023jwst} reveals dozens of point-source GC candidates in the VV~191 system not before seen with deep HST data. Throughout we use AB magnitudes ($m_{\rm AB}$) \citep{oke1983secondary} and the Planck flat $\Lambda$CDM cosmology with $H_0 = 70 \pm 0.3$ $km s^{-1}Mpc^{-1}$, $\Omega_\Lambda =$ 0.685, and $\Omega_m =$ 0.315 $\pm$ 0.007 \citep{collaboration2020planck, riess2022comprehensive}. With the above cosmological values, we find the luminosity distance of our target galaxy to be 228 $\pm$ 10 Mpc, using a redshift of 0.0513 \citep{keel2023jwst}, and the Wright Cosmology Calculator \citep{wright2006cosmology} (See Appendix A for justification.) Section 2 of this paper describes the observations and data reduction; Section 3 gives the color--magnitude data which shows a fairly uniform population of GCs; Section 4 discusses the results of the cluster uni.- vs. bimodality; Section 5 discusses the GC stellar evolution modeling, followed by the LF in Section 6, and ending with our conclusions in Section 7.

\section{Observations and Data Reduction}
The following section discusses the JWST data, point spread function (PSF) matching, photometry, the Vega to AB transformation needed for model comparison, and a discussion of sample contamination, reliability, and completeness.

\subsection{Data}

 The JWST "Prime Extragalactic Areas for Reionization and Lensing Science" (PEARLS) project has started to unveil the astonishing capabilities of the James Webb Space Telescope (JWST)\null. PEARLS is a GTO program (PID 1176; PI: R.~Windhorst) designed to provide the community with medium-deep imaging in up to eight near-infrared filters. PEARLS' main focus addresses two themes: First Light and Reionization and the Assembly of Galaxies \citep{windhorst23jwst}. PEARLS has also observed VV~191, an overlapping galaxy pair selected initially to provide a benchmark dust-attenuation profile for studying higher redshift, dusty environments. The galaxy pair consists of VV~191b --- a spiral galaxy in the foreground, and VV~191a --- an elliptical galaxy in the background situated at $z = 0.0513$ \citep{keel2023jwst}.

Previous imaging of VV~191 taken by the Hubble Space Telescope (HST) is too shallow to detect GCs. The JWST NIRCam data goes much deeper, and enables the detection of the older population of stars in the near-IR that make up many of these GCs. JWST imaging of VV~191 was obtained on 2 July 2022 as part of the PEARLS GTO program. The total exposures were 901 seconds in each F090W/F356W and F150W/F444W filter pair, all using MEDIUM8 readouts with three groups and three dithers, respectively, as detailed by \citep{keel2023jwst}. 

All detectors at each wavelength were drizzle-combined into single mosaic images. However, the analysis of VV~191 only uses the single short-wavelength (SW) detector encompassing the galaxies and matching cropped regions on the long-wavelength detectors. We refer the reader to \citet{keel2023jwst} for detailed steps in the data reduction process.

Visually, we observe what appears to be several dozen GCs surrounding the elliptical. Infrared observations taken by NIRCam provide a means to directly constrain the metallicity of GCs, in particular, the red GCs that are preferentially located in the high surface brightness inner regions of ellipticals \citep{kundu2007bimodal}. We will quantify these observations in the following sections. 

\subsection{Point Spread Function (PSF) matching}

Before accurate measurements can be made between the different wavelength NIRCam filters, we must note that images with different filters have different PSFs. The PSF Full Width at Half Maximum (FWHM) values in each band for the VV~191 system are found in \citet[][see their Table 1]{windhorst23jwst}. PSF matching is achieved by convolving the image with a kernel generated from the PSF corresponding to the image and the reference PSF \citep{boucaud2016convolution}. To accurately conduct multi-band photometry on the globular clusters in VV~191, we first convolved all images with a kernel to match the filter with the broadest PSF, which in our case is F444W.

PSFs were created with WebbPSF \citep{perrin2012simulating} with a wavefront measurement done on July 2, 2022, i.e., $\sim$1.07 days before observations were taken. To convolve the short wavelength images to the longer wavelength PSF, we need a kernel that is the ratio of the two PSFs. However, doing this on the WebbPSFs introduces artifacts. To address this, we smooth the noise with windowing functions: tophat, cosine bell, and split cosine bell windows. These windowing functions are then multiplied by the Fourier transform of the PSF. The PSFs and windowing functions were grid-searched to find the windowing function that best reproduced the F444W PSF along a 1D profile through the centers of the PSF. The grid search was done as follows: (1) starting at parameter values of 0.01, each window was used to convolve the F090W PSF with the parameter value passed to each of the windowing functions; (2) a 1D profile from the center of the convolved PSF was compared to the 1D-profile from the source F444W PSF along the same pixel values using a chi-square routine; (3) the optimal chi-square value and windowing function were recorded as the combination to use for each of the images. The split cosine bell windowing function appeared optimal in all cases, with an $\alpha$ value of 0.4 and a $\beta$ value of $-0.1$. The $\alpha$ value is the percentage of tapered array values, and the $\beta$ value is the inner diameter as a fraction of the array size beyond which the taper begins and must be $\leq$ 1.0.

We note that the radial variation of the PSF across the VV~191a elliptical galaxy is rather small since the galaxy does not span more than 800 pixels across compared to a total of the 4096$^2$ drizzled pixel FOV of each module. The GC photometric apertures were chosen wide enough to encapsulate the impact of field- and azimuth-dependent PSF asymmetry on the total magnitudes, which is small in any case since our GC candidates are so faint.

\noindent\begin{figure*}[ht]

\centering
\includegraphics[width=\textwidth]{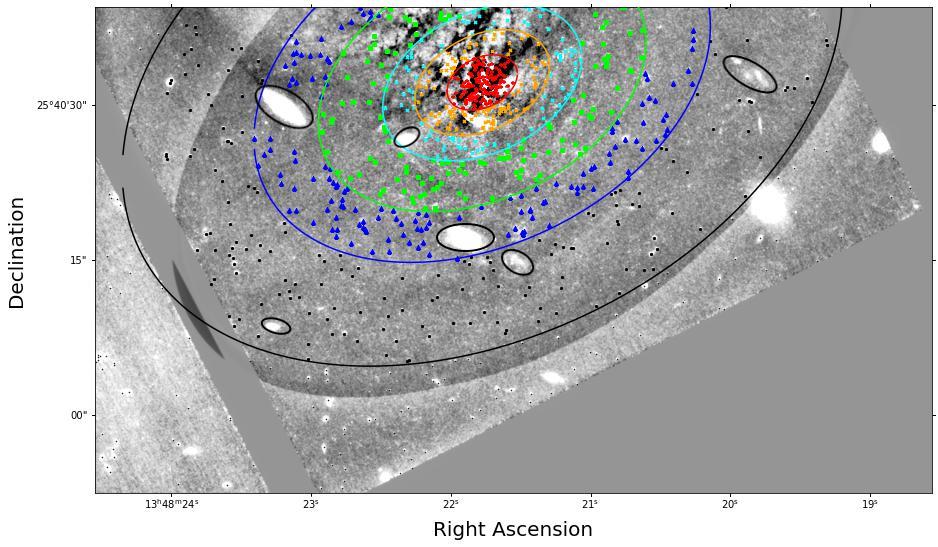}
\caption{The VV~191a image is shown after calibration and galaxy subtraction were performed. We used {\sc Photutils} to map out point source aperture objects in the F090W filter using a 3-pixel aperture radius. We further subdivide the region usable for GC searches --- i.e., away from the outer spiral arm of the foreground spiral --- into six elliptical sections. With this method, we could compare the colors of GC candidates within the galaxy using CMDs from data in each ellipse. Further details are given in section 2.5 \& 3, as well as Fig. \ref{fig:f090_f150_CMD_Hist} \& \ref{fig:f356_f444_CMD_Hist}. The black ellipses are bright background galaxies, and their point sources are not included in the analysis.}
\label{fig:starfinder}
\end{figure*}

\subsection{Photometry}

The main advantage of aperture photometry is that it provides a simple measure of the total light within a fixed radius, and is less affected by small object-to-object differences in the actual profile shape. All photometry was performed on the galaxy-subtracted image (see \cite{keel2023jwst} for details). Before starting the analysis, the image is cropped to exclude the spiral galaxy as much as possible, further reducing contamination from stars and dust found within and around the spiral arms. The objects outlined by the small black ellipses are also excluded (see Fig. \ref{fig:starfinder}) and recognized as diffuse objects (i.e., bright background/foreground galaxies). Using the {\sc Photutils} \citep{larry_bradley_2022_6825092} package DAOStarFinder, positions of point-source objects are located using the NIRCam F090W filter as the reference image. The Python class DAOStarFinder is a routine that implements the DAOFIND algorithm \citep{stetson1987daophot}.  It searches a {\it convolved} image for local density maxima with a peak amplitude greater than a threshold of 0.044 (in MJy/sr or $\sim$10$\times$ the F090W rms image noise) and a size and shape similar to a defined 2D Gaussian kernel FWHM of $\simeq {4.7}$ pixels, closely matching the F444W PSF. This approach works particularly well since virtually all the individual GCs appear starlike since our target galaxy has a luminosity distance $(d_L)$ of approximately  228 $\pm$ 10 Mpc. Point sources within the F090W filter are initially stored as $x$ and $y$ pixel coordinates before being converted to celestial coordinates. The celestial coordinates of the point sources found in the F090W filter will be the reference coordinates used to locate the corresponding sources in the F150W, F356W, and F444W filters and are referred to as ``aperture objects''. Once aperture objects have been created, we can perform photometry on the input data by summing the flux within the given aperture. The apparent magnitude is defined as:
 \begin{equation}
     m = -2.5\ \log_{10}(\phi)\ +\ {\rm ZP}.
 \end{equation}
 with the flux-density measurements, $\phi$, from each individual point source object detected by the DOAStarFinder algorithm and zero point $\rm (ZP) = 28.0865$ mag for a pixel scale of 0\arcspt 030 per pixel \citep{windhorst23jwst}. We can then compute the absolute magnitude by determining the distance modulus $(m-M) = 36.79$ mag from:
 \begin{equation}
     m-M = 5\ \log_{10}\left(\frac{d_L}{\rm Mpc}\right)+25\ mag.
 \end{equation}

\subsection{AB to Vega transformation need for model comparison}

Unlike ground-based or HST observations, there was a non-obvious redefinition of the JWST Zeropoints (ZPs), which also affects the Vega- to AB-magnitude conversion needed for the model comparisons. The AB-magnitude zeropoint for JWST images is defined in units of MJy/sr and therefore depends on the pixel scale of the mosaics used. The details of this are described in section 3.3 and Eq. 2 of \citet{windhorst23jwst}. As a consequence, the conversion of model Vega to observed AB-magnitudes is no longer a single-wavelength dependent constant, as is typical for all HST and ground-based data, but needs an extra term, as described on the STScI website\footnote[1]{\url{https://jwst-docs.stsci.edu/jwst-near-infrared-camera/nircam-performance/nircam-absolute-flux-calibration-and-zeropoints}}. We thus use the additional constants $PHOTMJSR$ as provided in the STScI table\footnote[2]{\url{https://jwst-docs.stsci.edu/files/182256933/224166043/1/1695068757137/NRC_ZPs_1126pmap.txt}}. 

\begin{deluxetable*}{lccccccc}
\tablecaption{GC candidates and background galaxy contamination } 
\label{f090_ringlet_table}
\setlength{\tabcolsep}{7pt}
\tablehead{
\colhead{Ringlet} & \colhead{Ringlet Area} & \colhead{Number of GC} &  \colhead{Num of contaminating} & \colhead{Net number of } & Sample & Corrected Total \\
\colhead{Nr.}     & \colhead{(arcsec$^2$)} & \colhead{candidates}   &  \colhead{background galaxies}  & \colhead{reliable GCs$^1$}   & Reliability  & Number of GCs \\
\colhead{(1)} & \colhead{(2)} & \colhead{(3)} & \colhead{(4)} & \colhead{(5)} & \colhead{(6)} & \colhead{(7)} \\
}
\startdata 
1      &   27.13 &  52 &  1.0 &   51.0 & 98.0 \%  & 51.0 \\ 
2      &   72.55 &  38 &  2.8 &   35.2 & 92.7 \%  & 35.2 \\
3      &  118.19 &  25 &  4.5 &   20.5 & 82.0 \%  & 20.5 \\
4      &  293.99 &  37 & 11.1 &   25.9 & 69.9 \%  & 32.9 $^2$ \\
5      &  373.78 &  36 & 14.3 &   21.7 & 60.3 \%  & 32.3 $^2$ \\
\hline
SubTot &  885.64 & 188 & 33.73 & 154.3 & 82.1 \% & 171.9\\
\hline
6      & 1042.95 &  55 & 39.84 &  15.2 & 27.6 \% & \\
\hline
Total  & 1928.59 & 243 & 73.57 & 169.5 & 69.7 \% & \\
\enddata	 
\sn {\it Notes to Table:}\ 

\n $^1$ The net number of reliable GCs is the observed number of GC candidates minus the number of background galaxies predicted for each ringlet area adopting the galaxy counts of \citet{windhorst23jwst}. 

\n $^2$ The outer two elliptical ringlets 4--5 used in our sample were cut off in Fig. 1, and the corrected total number of GCs listed here includes the fraction missed in the cut-off area of Fig. 1, assuming elliptical symmetry.
\end{deluxetable*}

\begin{figure}[ht]
\centering
\includegraphics[width=0.5\textwidth]{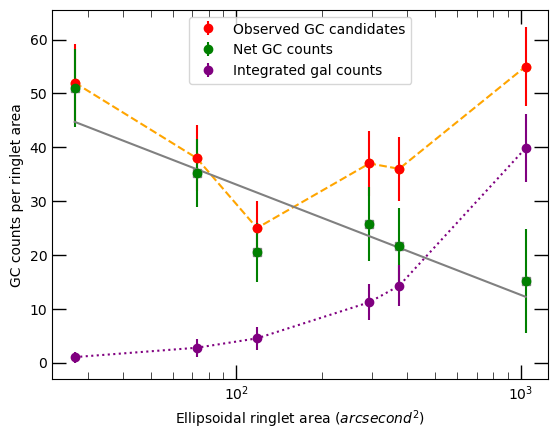}
\caption{Plot showing the number of observed GC candidates (orange dashed line), the number of contaminating BG galaxies (purple dotted line), and the net number of reliable globular clusters (green) as a function of ringlet area. It is clear that the first five ringlets show a significant number of real GCs, but contaminating background galaxies dominate the sixth ringlet. The net number of GCs that we trust (green) is a declining function of radius or area.}
\label{fig:v-shape}
\end{figure}

\subsection{Sample Contamination, Reliability, and Completeness}

As discussed in Section 5, the elliptical galaxy is mainly transparent at near-IR wavelengths. Therefore, a number of background galaxies may have leaked into our GC sample. To a good approximation at $z = 0.0513$, GCs are point sources to JWST NIRCam, but so are many faint galaxies, as discussed in Section 4.5 of \citet{windhorst23jwst}, especially at the shallow and minimally sampled images of VV~191. With only 4 NIRCam filters available for the original science goals of VV~191, it is not possible to make accurate photometric redshift estimates for background galaxies behind VV~191a. However, we can make a {\it statistical correction} for the number of background galaxies behind VV~191a in each elliptical ringlet of Fig. \ref{fig:starfinder} in our selected flux range of 24.0\cle $m_{\rm AB}$\cle 28.65 mag. While other sources of contamination (e.g. supergiants in the foreground spiral) may be present, the fact that this approximation results in a smoothly declining GC number with radius (Fig.~\ref{fig:v-shape}) suggests that it is sufficient for this analysis.

Table \ref{f090_ringlet_table} and Fig. \ref{fig:v-shape} show our procedure to estimate the number of background galaxies behind VV~191a, and how these may affect the GC sample in each of the ringlets in Fig. \ref{fig:starfinder}. Table \ref{f090_ringlet_table} shows each of the adopted elliptical ringlet areas in $arcsec^2$, the number of detected GC candidates, the number of background galaxies from $m_{\rm AB}$ $=$ 24 mag down to the detection limits (AB\cle 27.2--28 mags) expected in each ringlet area, following Fig. \ref{fig:col_hist}--8 and section 4 of \citet{windhorst23jwst} and the integration method of \citet{tompkins2023cosmic}, with a cosmic variance on the integrated galaxy counts of $\le$10\% \citep[see section 4 and Appendix B2 of][]{windhorst23jwst}. This number amounts to 496,300 galaxies deg$^{-2}$ (or 0.0383 arcsec$^{-2}$) over the magnitude range 24\cle $m_{\rm AB}$\cle 27.2 mag in F090W. We note that the bulk of the faint background galaxies is small enough \citep{windhorst23jwst} in the very shallow, poorly sampled VV~191 images to be identified as point sources by DAOStarfind.
Column 4 of Table~\ref{f090_ringlet_table} lists the resulting number of expected interloping galaxies for each of our elliptical ringlet areas, and column 5 shows the {\it net} number of reliable globular clusters remaining in the VV~191a sample. Column 6 then lists the resulting sample reliability (\ie\ true GC abundance over total observed objects). For ringlets 4--5, column 7 in Table~\ref{f090_ringlet_table} corrects the reliable number of GCs found for the fact that their geometric area was cut off in Fig.~\ref{fig:vv191} (to avoid the foreground spiral). This correction assumes elliptical symmetry in the overall GC distribution. 

The results are plotted in Fig. \ref{fig:v-shape}. The uncertainties in the contaminant abundance come from the 10\% cosmic variance described above; the uncertainties in the GC candidate abundance are Poisson errors, and the uncertainties in the net GC counts come from the combination of the two. It is clear from Fig.~\ref{fig:v-shape} that for the inner 1--5 ringlets, the number of expected background galaxies shining through the elliptical is relatively small, and generally much smaller than the number of actual GCs, but this is no longer true for ringlet 6, which we therefore display, but {\it do not} include in our final GC sample. Fig.~\ref{fig:v-shape} shows that our net number of reliable GCs declines with radius or area samples, as is expected in studies of more nearby elliptical such as M87 \citep[\eg][]{forbes2012baryonic}. 

We thus adopt the inner five ringlets as our final GC sample, with the contamination subtraction as in Table 1 and Fig. \ref{fig:v-shape}. We emphasize that because we only have 4 NIRCam filters for V191a, we cannot actually identify {\it which GC candidates} are background galaxies, and so we can only make a statistical correction for these contaminants, which we propagated throughout the paper. (A more detailed discussion of the color distribution of the BG galaxies compared to that of our GC candidates is given in Appendix B.) In summary, Table 1 shows that the overall sample of GC candidates is 188 in size, with an expected number of $\sim$34 contaminating background galaxies to $m_{\rm AB}$\cle 27.2 mag within the 886 arcsec$^2$ of the outermost used ringlet 5, and therefore we have $\sim$154 plausible GC candidates to $m_{\rm AB}$\cle 27.2 mag. This number is in line with the expectations, down to our limiting magnitude, for an elliptical with a total mass of $\sim$9$\times$$10^{11}$ $M_{\odot}$ \citep[\eg][]{saulder2016matter, diego2023jwsts}, given the uncertainties in the mass estimates. 

Given the statistical contamination of field galaxies described above in Table~\ref{f090_ringlet_table} and Section 4, we also need to address the (F090W--F150W) and (F356W--F444W) color-distributions of field galaxies above our completeness limits and the number of contaminating objects they may cause in the wings of our GC color distributions. To assess this, we will use the fully reduced and cataloged data from our PEARLS 48-hour North Ecliptic Pole (NEP) Time Domain Field (TDF), which provides NIRCam 8-filter galaxy catalogs to $m_{\rm AB}$\cle 28.5--29.2 mags \citep{windhorst23jwst}.

\begin{figure*}[ht]
\centering

\includegraphics[width=\textwidth]{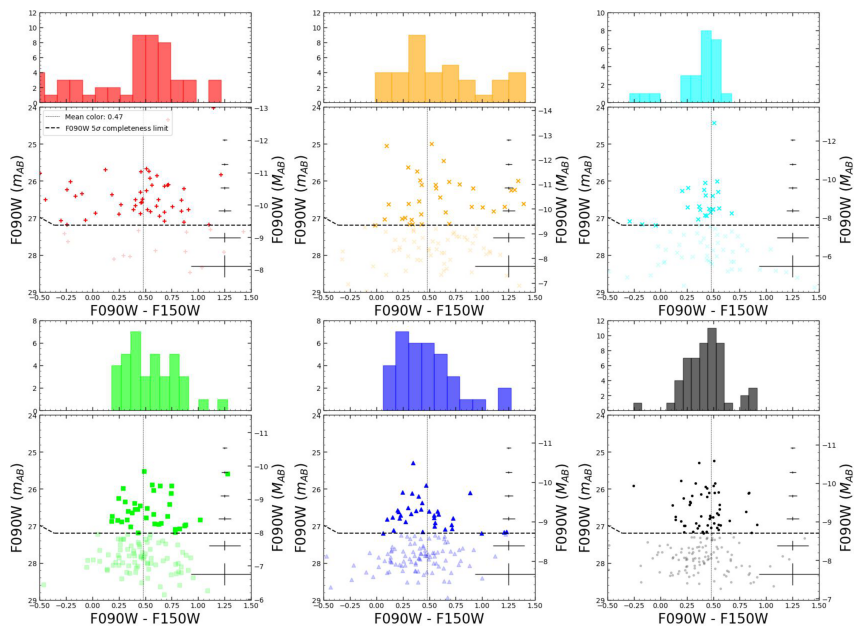}
\caption{CMDs of F090W magnitude versus the (F090W--F150W) color for our GC sample. 
\DELETED{Each of the six panels corresponds to a specific ellipse in Fig. \ref{fig:starfinder},}
Histograms are color-coded to match the colors of the elliptical regions depicted in Fig. 2, with the top-left panel representing data from the center ellipse and then moving outward. In all six panels of the CMDs, we see a distinct population of GCs with a mean color of 0.47 mag and a standard error of 0.029 mag.}
\label{fig:f090_f150_CMD_Hist}
\end{figure*}

\begin{figure*}[ht]
\centering

\includegraphics[width=\textwidth]{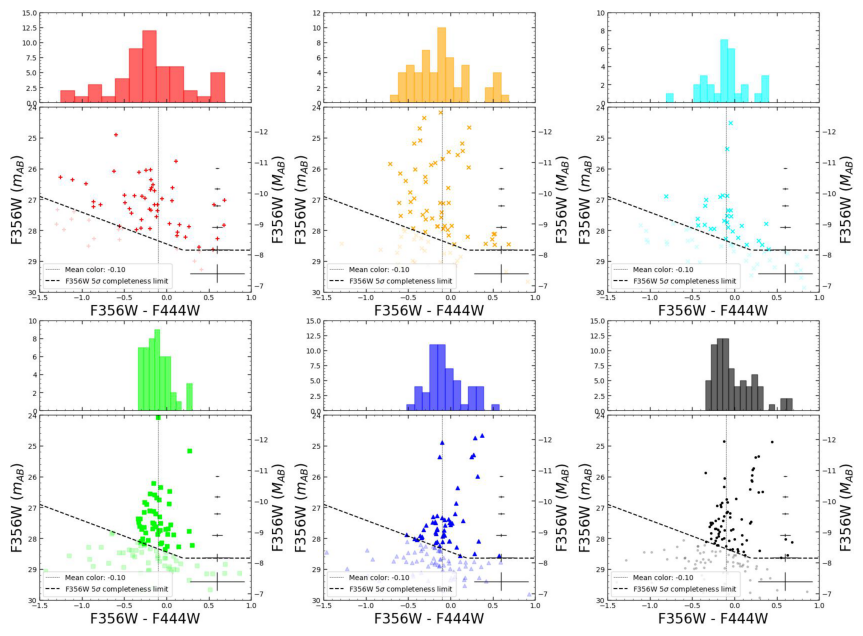}
\caption{CMDs of F356W magnitude versus the (F356W--F444W) color for our GC sample. 
\DELETED{Each of the six panels corresponds to a specific elliptical ringlet in Fig. \ref{fig:starfinder}} 
Histograms are color-coded to match the colors of the elliptical regions depicted in Fig. 2, with the top-left panel representing data from the center ellipse and then moving outward. In all six panels of the CMDs, we see a distinct population of GCs with a mean color of -0.103 mag with a standard error of 0.019 mag.}
\label{fig:f356_f444_CMD_Hist}
\end{figure*}

We compare the resulting field galaxy color distributions to our CG color distribution in Figs.~\ref{fig:f090_f150_CMD_Hist}, \ref{fig:f356_f444_CMD_Hist}, and \ref{fig:GC_models}ab. To first order, the (F090W--F150W) and (F356W--F444W) color distributions of our conservative GC samples with F090W\cle 27.2 mag in Table~\ref{f090_ringlet_table} are fit by a single main Gaussian in Fig. \ref{fig:GC_models}ab, but some additional red and blue GC candidates can be seen in both histograms. Section 4 further discusses these color outliers.


\section{The Color--Magnitude Data}

 We identify GC candidates across most of the elliptical galaxy, a fraction of which is obscured in the foreground by the dust associated with VV~191b. Fig. \ref{fig:f090_f150_CMD_Hist} and \ref{fig:f356_f444_CMD_Hist} show the color--magnitude diagrams. Each of the six panels corresponds to a specific ellipse in Fig. \ref{fig:starfinder}, with the top-left panel representing data from the center ellipse, moving outward. We chose to analyze the data in sections to check that the foreground spiral was not adding significant reddening to the GC sample, and to study any possible variation of GC properties with a galactocentric radius. We conclude this is not the case when we look at the color distribution in each of the six panels. We would expect the objects closer to the spiral to have a redder color distribution if there was some contamination from spiral arm dust. However, this does not appear to be the case in our observations. 
 
Furthermore, dust extinction in elliptical galaxies is generally rather small. A previous study by \cite{white2000seeing} investigated dust extinction in the overlapping galaxy-galaxy pairs, AM 0247-312 (S0 + E), IC 5328 (E + S0), and UGC 8813 (S0 + S0), and found that in all cases that foreground E and S0 galaxies show no measurable absorption: $A_{B} \leq 0.1$ mag in their B-band observations, which would imply \AV\cle 0.076 mag for a standard Galactic extinction curve. \cite{bonatto2013mapping} found extinction in 66 GCs with mean values ranging from $<\delta E(B -v)> \approx 0.018$ (NGC 6981) to $<\delta E(B-v)> \approx 0.016$ (Palomar 2), and therefore an implied \AV$\sim$ 0.05 mag, consistent with the above value. More recent work by \citet{kim2019analysis} made a pixel-pixel analysis of a large sample of nearby ellipticals with ground-based and Spitzer 3.6 \mum\ images, and found that the radial extinction profiles in the elliptical centers have on average \AV\cle 0.1 mag, and generally do not exceed \AV $=$ 0.2 mag in their central reddening values.

We will thus adopt \AV $=$ 0.1 mag as a working hypothesis for VV~191a's extinction value
in the visual. The absorption is considerably less in the near-IR since extinction scales approximately as $\lambda^{-1}$. For a standard Galactic extinction curve, this then results in the following adopted limits to the reddening values in VV~191a: A$_{F090W}$\cle 
0.062 mag, A$_{F150W}$\cle 0.037 mag, A$_{F356W}$\cle 0.016 mag, and A$_{F444W}$\cle 0.012 mag \citep{cardelli89gal}. Their corresponding reddening vectors are plotted in Fig. \ref{fig:GC_models}ab for a unit reddening of A$_{\lambda}\equiv$1.0 mag, \ie\ much larger than the actual reddening values. The typical reddening values above in elliptical galaxies are 16$\times$ smaller than the plotted reddening vector in Fig. \ref{fig:GC_models}a, and $\sim$60$\times$ smaller than the reddening vector in Fig. \ref{fig:GC_models}b. Hence, any discrepancy between the data and the models in Fig. \ref{fig:GC_models}ab cannot be readily explained by reddening.

Objects plotted below the bold dashed line fall below the 5$\sigma$ point source completeness limit \citep[][]{windhorst23jwst}, and, therefore, are not formally part of a complete sample of GCs. Data analysis and figures going forward will only include objects detected above the 5$\sigma$ completeness limit. The 5$\sigma$ completeness limits of each filter are shown in Table \ref{tbl-completeness lim}. We find a mean (F090W--F150W) color of 0.476 mag with a standard error of 0.029 and a mean (F356W--F444W) color of --0.103 mag with a standard error of 0.019.


We attach histograms to the 6-panel plots in Figs. \ref{fig:f090_f150_CMD_Hist} and \ref{fig:f356_f444_CMD_Hist} to better visualize the color distributions as a function of galactocentric radius. Fig. \ref{fig:f090_f150_CMD_Hist} and \ref{fig:f356_f444_CMD_Hist} plots the combined color-magnitude error above the 5$\sigma$ catalog completeness limits and has a reasonably small spread compared to the photometric uncertainties ($\sim0.02$ mag). Fig. \ref{fig:f090_f150_CMD_Hist}--\ref{fig:f356_f444_CMD_Hist} shows that the two innermost ringlets show the widest color spread, and may indicate some color bimodality in both the F090W vs. (F090W--F150W) and the F356W vs. (F356W--F444W) CMDs, though it is small. Hence, above our conservative completeness limits that we apply to all ringlets, the relatively small errors do not wipe out any bimodality much larger than those in the CMDs. Section 4 discusses this further.

\begin{deluxetable}{lcccl}
\tablecaption{5$\sigma$ Point-source AB Mag Limit}
\label{tbl-completeness lim}
\centering
\setlength{\linewidth}{7pt}

\tablehead{
\colhead{F090W} & \colhead{F150W} & \colhead{F356W} & \colhead{F444W} & \colhead{Source} \\}
\startdata
27.88 & 28.24 & 29.01 & 28.81 & \citep{windhorst23jwst} \\
27.20 & 27.56 & 28.65 & 28.45 & Adopted limits for VV~191center \\
\enddata

\end{deluxetable}

\begin{figure*}[ht]
\centering
\includegraphics[width=\textwidth]{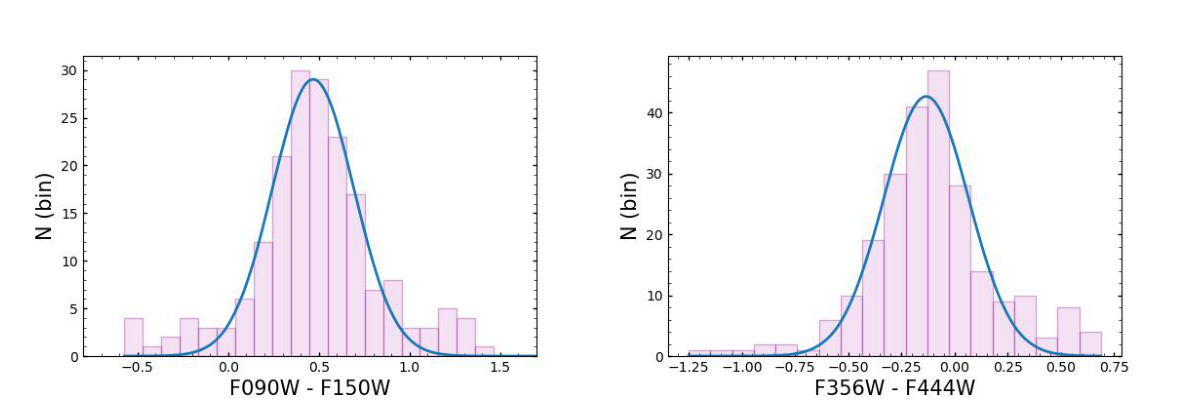}

\caption{Color histograms in (F090W--F150W) and (F356W--F444W) showing the fairly uniform color distribution of the VV~191a clusters. Fig. \ref{fig:col_hist}a shows 11 bluer objects (6\%) somewhat in excess of the single Gaussian fit with (F090W--F150W)\cle --0.2 mag, and 13 (7\%) redder objects in excess of the single Gaussian fit with (F090W--F150W)\cge 1.05 mag. Fig. \ref{fig:col_hist}b shows a total of 237 GC candidates in the F356W sample to $m_{\rm AB}$\cle 28.65 mag, Fig. b shows 13 bluer objects (5\%) somewhat in excess of the single Gaussian fit with (F090W--F150W)\cle --0.6 mag, and 24 (10\%) redder objects in excess of the single Gaussian fit with (F356W--F444W)\cge 0.3 mag.}

\label{fig:col_hist}
\end{figure*}
\section{GC Unimodality vs. Bimodality}

In many of the past studies of GC systems, authors have used color histograms and fit uni- vs. bimodal Gaussian to decide whether the system falls into a bimodal color distribution, indicating a population of metal-rich and metal-poor GCs \citep[\eg][]{gebhardt1999globular, harris2009globular, larsen2001properties}. Many studies have used more traditional optical colors, where the bimodality of GCs is more visible because of the stronger metallicity dependence of optical colors. Two recent studies by \cite{hartman2023comparing, harris2023photometric} present HST observations in many nearby galaxies and perform photometry of their GC systems. Both of these studies use optical/near-IR filters to constrain the metallicity distribution of GCs, and indeed, these optical color indices are typically more metallicity-sensitive \citep{harris2006globular}. In general, optical--near-IR GC colors depend on both metallicity and age. A study by \cite{kundu2001new} used HST to study GC bimodality using (V--I) color distributions and found that 17 of the 29 galaxies appear to have a bimodal GC population.

In contrast, a more recent study by \cite{harris2023jwst} uses JWST near-IR photometry to investigate the GC population of Abell 2744 at $z = 0.308$. In their study, they find that NIRCAM color indices in the (F115W--F150W) and (F150W--F200W) filters are not strongly dependent on metallicity. Instead, they found that the CMDs in these colors form a single narrow, mostly vertical sequence, showing no evidence of bimodality in the (F115W--F150W) color, and very minimal bimodality in the (F150W--F200W) color. They thus find that the color with the widest wavelength baseline (F115W--F200W) is about twice as metallicity-sensitive as either of these other colors. As a consequence, their GC color distribution is distinctly broader in this filter combination.

Fig. \ref{fig:col_hist}ab shows color histograms to determine the color distribution of the clusters and see if these show evidence for bimodality. Since most of the background contamination is found in the outer-most ringlet (see Table \ref{f090_ringlet_table}), we only use data from ringlets 1--5 to test the color distribution of the sample. For a total of 188 GC candidates in the F090W sample to $m_{\rm AB}$\cle 27.2 mag, Fig. \ref{fig:col_hist}a shows 11 bluer objects (6\%) somewhat in excess of the single Gaussian fit with (F090W--F150W)\cle --0.2 mag, and 13 (7\%) redder objects in excess of the single Gaussian fit with (F090W--F150W)\cge 1.05 mag. The number of NEP TDF field galaxies within the {\it same total ringlet area} as our GCs in Table \ref{f090_ringlet_table} amounts to $\sim$3 interloping blue objects with (F090W--F150W)\cle --0.2 mag, so that there may be $\sim$8 real GCs (4\%) in the blue wing of the (F090W--F150W) color distribution in Fig. \ref{fig:col_hist}a. Similarly, there are $\sim$4 interloping red background galaxies with (F090W--F150W)\cge 1.05 mag in the total area of ringlets 1--5, so there may be $\sim$9 (5\%) real GCs in the reddest wing of the (F090W--F150W) color distribution in Fig. \ref{fig:col_hist}b. 

In the (F356W--F444W) color, a similar comparison holds but is applied to its deeper detection limits. For a total of 237 GC candidates in our complete F356W sample to $m_{\rm AB}$\cle 28.65 mag, Fig. \ref{fig:col_hist}b shows 13 bluer objects (5\%) somewhat in excess of the single Gaussian fit with (F090W--F150W)\cle --0.6 mag, and 24 (10\%) redder objects in excess of the single Gaussian fit with (F356W--F444W)\cge 0.3 mag. The number of NEP TDF field galaxies within the {\it same total ringlet area} as our GCs in Table 1 amounts to 9 interloping blue objects with (F356W--F444W)\cle --0.6 mag, so that there may be $\sim$5 (2\%) real GCs in the blue wing of the (F356W--F444W) color distribution in Fig. \ref{fig:col_hist}b. Similarly, there are $\sim$6 interloping red background galaxies with (F356W--F444W)\cge 0.3 mag in the total area of ringlets 1--5, so there may be $\sim$18 (8\%) real GCs in the red wing of the (F356W--F444W) color distribution in Fig. \ref{fig:col_hist}b. In both colors, the fraction of very red objects is somewhat larger than the fraction of very blue objects, and this could well be real.

\begin{figure*}[ht]
\centering
\includegraphics[width=\textwidth]{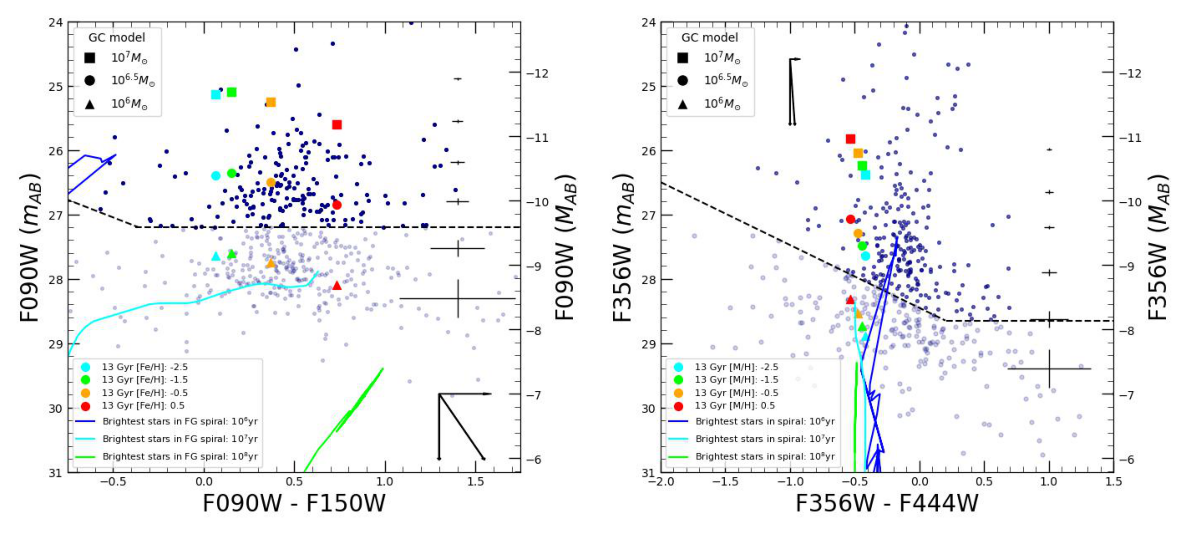}
\caption{CMDs showing GC models in the F090W vs (F090W--F150W) and F356W vs (F356W--F444W) filters. The larger triangles represent the model which describes a GC mass of $10^6 M_{\odot}$,  the circle represents the model using $10^{6.5} M_{\odot}$, and the square represents the model using $10^{7} M_{\odot}$. The solid lines are stellar isochrone models for the (hypothetical) brightest individual stars in the foreground spiral VV~191b at $z = 0.0514$, showing that objects detected in the CMDs are mostly GCs in VV~191a, and in general, not consistent with luminous stars in the foreground spiral. The diagonal dashed line is the 5$\sigma$ completeness limit. The black vectors (in the bottom right and top left) represent the effect of reddening of $A_{\lambda} =$ 1.0 mag (much higher than we expect for our GCs) on the data.}
\label{fig:GC_models}
\end{figure*}

In summary, after correcting for the major known sources of (color-dependent) contamination --- the bluest or reddest background field galaxies --- the (F090W--F150W) and (F356W--F444W) color distributions appear to be largely single Gaussians, but with some excess very blue component that comprises 2--4\% of the total objects, and a somewhat larger excess very red component that comprises 5--8\% of the total number of GCs. We briefly mention some possible explanations of these. Fig. \ref{fig:GC_models}a shows that there could be some interloping very young blue stars, although we argue below that these likely could not have dispersed into the very outskirts of the spiral VV~191b in front of our elliptical VV~191a during their stellar lifetimes.

\noindent
 
\section{GC and stellar modeling}

\begin{figure}[h]
\centering
\includegraphics[width=0.5\textwidth]{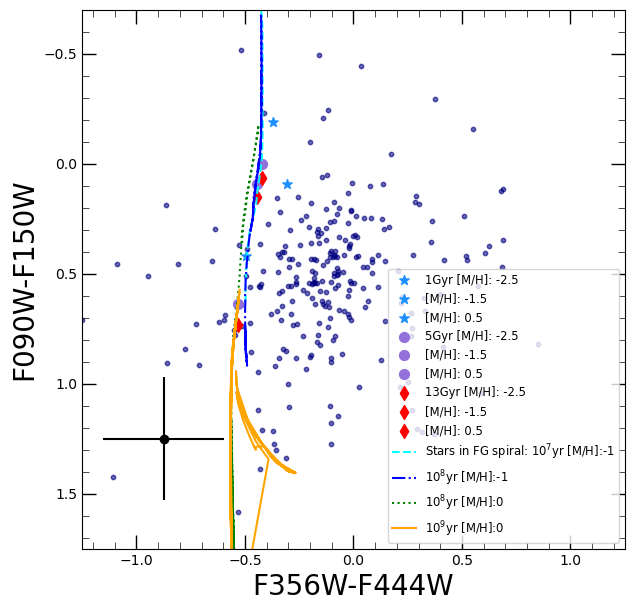}

\caption{Color-color diagram plotted with both stellar isochrone and GC models. The GC model represents ages of 1 Gyr, 5 Gyr, and 13 Gyr, while the vertical turquoise-blue-green-orange) dot-dashed isochrone model represents various stellar ages and metallicities of the brightest stars that may be found within the spiral galaxy. These models lie askew of the area populated by the data. However, we find that the GC models also do not represent the data accurately, with the GC data generally being redder than the models in the reddest NIRCam filters.}
\label{fig:star_model}
\end{figure}

When modeling the GC population, we only use data from ringlets 1--5 to avoid excess contamination from the outermost ring. The CMDs in Fig. \ref{fig:GC_models}ab, show F090W vs (F090W--F150W) and F356W vs (F356W--F444W). We use the PARSEC CMD 3.7 model using single-burst stellar populations, integrated magnitudes (for 1 $M_\odot$) \citep{salpeter1955luminosity, chabrier2003galactic} with a fixed age of 13 Gyr while varying the GC mass using $10^6 M_{\odot}$ represented by a triangle, $10^{6.5} M_{\odot}$ by a circle, and a square representing the model that uses $10^{7} M_{\odot}$, to address how well the model fits our data. This is shown in Fig. \ref{fig:GC_models}. We chose to use an age of 13 Gyr because when reducing the age of the model somewhat, there is very little change in the predicted colors. Fig. \ref{fig:star_model} shows this trend in more detail.

Fig. \ref{fig:GC_models}a and Fig. \ref{fig:GC_models}b show a distinct difference 
in the model--data comparison for the (F090W--F150W) and (F356W--F444W) colors. The models in the (F090W--F150W) color fit the data quite well. These models show that a majority of the GC have masses of $10^{6.5}  M_{\odot}$, with metallicities [Fe/H] spanning the typical range expected of GCs, from $\simeq -2.5$ to $0.5$. However, the observed GCs in the (F356W--F444W) colors are, on average, $\sim$+0.3--0.4 mag redder than the models for a reasonable range of metallicities. 

To check if reddening or extinction in the GCs themselves could be causing discrepancies in the models, we have run the models for several $A_V$ values (0.1, 0.2, 0.3) and added the $A_\lambda$ reddening vectors to Fig. \ref{fig:GC_models}. To first order, assuming that the GC extinction is not much larger than that in the elliptical itself, this range of $A_V$ values assumed does not change the model locations compared to the data in a significant fashion. This is consistent with the small anticipated A$_{\lambda}$ values in elliptical galaxies discussed in section 3. We currently do not have a good explanation for this discrepancy between the (F356W--F444W) models and data, and for that reason, we cannot identify a cause for either the majority of the (F356W--F444W) colors or for the 5--8\% fraction of very red ones in Fig. \ref{fig:col_hist}. For any reasonable A$_{\lambda}$-values, the reddening vectors in Fig. \ref{fig:f090_f150_CMD_Hist}ab cannot explain this discrepancy either, because they mostly run vertically in Fig.7b and the main (F356W--F444W) data-model offset is horizontal. 

This discrepancy also cannot be explained by the youngest stars in the foreground spiral, as these would take too long to diffuse well outside the dust in the spiral arms where they formed, as mapped by \citet{keel2023jwst}. Given that the same models --- which include standard prescriptions for AGB stars --- readily explain the observed (F090W--F150W) color distribution in Fig. \ref{fig:GC_models}a, it is not clear what causes the +0.3--0.4 mag model--data discrepancy in (F356W--F444W). Dust production of AGB stars in GCs may need to be reconsidered. Given this uncertainty, both the (F090W--F150W) and the (F356W--F444W) models suggest that the bulk of the GCs above our detection limits have masses of order 10$^{6.5}$ \Mo. In (F090W--F150W), the fraction of GCs with implied masses $\sim$10$^7 M_{\odot}$ is very small, while in (F356W--F444W), it is larger, but we cannot address this in more detail until the (F356W--F444W) model--data discrepancy is understood. Addressing these aspects will require much deeper HST UV--optical and deeper JWST filters and images, which have been or will be proposed for and will be the focus of future papers.

Fig. \ref{fig:GC_models}a and b shows that all young stars in the foreground spiral with ages $\sim$10$^7$--10$^8$ years are much fainter than our conservative detection limits. Even the youngest stellar models with ages 10$^6$ years do not explain more than 5--6 of the bluest GC candidates in the F090W vs. (F090W--F150W) CMD of Fig. \ref{fig:GC_models}a. However, in the F356W vs. (F356W--F444W) CMD of Fig. \ref{fig:GC_models}b, the 10$^6$ yr age models do indeed penetrate $\sim$1 mag into the area of detected GC candidates. The polygon in Fig. \ref{fig:vv191}, over which we search for GCs, excludes, by definition, most of the visible foreground spiral VV~191b, and while there may be a few such stars that exist, we do not expect the number of bright stars from the foreground spiral to be large in this region. This is simply because the southernmost dim spiral arm with measurable dust in VV~191b is four kpc north of the core of the elliptical VV~191a at their respective distance using our cosmology \citep[see Fig. 1 of][]{keel2023jwst}. Most of our GC candidates in the selection region in Fig. \ref{fig:vv191} are further south of there. Hence, the closest supply region of bright stars in the foreground spiral is at least five kpc from where we find most of our GC candidates in VV~191a.
At the typical diffusion speeds of young stars from their birth environment of tens of km/sec at most, it would have taken \cge 200 Myrs for such foreground stars to significantly pollute our sample in the sample selection regions of Fig. 1--2. Indeed, we see \cle 5 possible bright $10^6$ yr old star candidates in Fig. \ref{fig:GC_models}a -- well above the detection limit. We do not expect this number to be much larger in Fig. \ref{fig:GC_models}b for the above dynamical reasons.

In detail, we note that Fig. \ref{fig:GC_models}a has 188 GC candidates above the completeness limit, and Fig. \ref{fig:GC_models}b has 237 GC candidates. A significant fraction of this $\sim$52 GC excess in F356W/F444W is likely due to the deeper detection limits in the NIRCam LW modules, following the discussion in section 4 of \citet{windhorst23jwst}. To a good approximation, the shape of the radial dependent (F090W--F150W) and (F356W--F444W) CMDs in Fig. \ref{fig:f090_f150_CMD_Hist}--\ref{fig:f356_f444_CMD_Hist} are very similar, and the outer ringlet is well away ($>>$ 5 kpc) from the foreground spiral, so we will henceforth assume that the majority of objects detected in the CMDs of Fig. \ref{fig:GC_models}ab are not likely stars from the foreground spiral, but predominantly GC candidates in VV~191a.

For a final check of objects found within the respective completeness limit and assumed to be GCs, we use a color-color diagram plotted with PARSEC CMD 3.7 stellar isochrone model \citep{bressan2012parsec} as shown in Fig. \ref{fig:star_model}, along with GC models for reference. The stellar isochrone model is shown for stellar ages of $10^7$ -- $10^9$ years with a metallicity [M/H] of 0 and --1. The GC model covers ages of 1 Gyr, 5 Gyr, and 13 Gyr with varying metallicity [M/H]. As mentioned for the CMDs in the previous section, the current GC models (colored asterisks, circles, and diamonds in Fig. \ref{fig:star_model}) explain the range in (F090W--F150W) color well but are still $\sim$0.3--0.4 mag too blue in the (F356W--F444W) color compared to the JWST data. We can conclude that the current models need to be revisited to better represent the (F356W--F444W) colors.

\section{GC Luminosity function}

The LF, or the relative number of GCs as a function of magnitude, is one of the most fundamental properties of GC systems \citep{harris2014globular}. The turnover luminosity of the GCLF has been found to be exceptionally constant over a large range of galaxies and various environments, so much so that it has been used as a distance indicator \citep[\eg][]{harris1991globular, whitmore1997globular, rejkuba2012globular}. Previous studies with HST data in the V \& I  bands have shown a general turnover luminosity of the GCLF at $M_V \approx -7.5$ \citep{harris2009globular,rejkuba2012globular, kundu2001new}. Previous studies have reported the dispersion, $\sigma$, of the Gaussian describing the GCLF as varying considerably from one galaxy to the next \citep{secker1993maximum}. 

\begin{figure}
\centering
\includegraphics[width=0.5\textwidth]{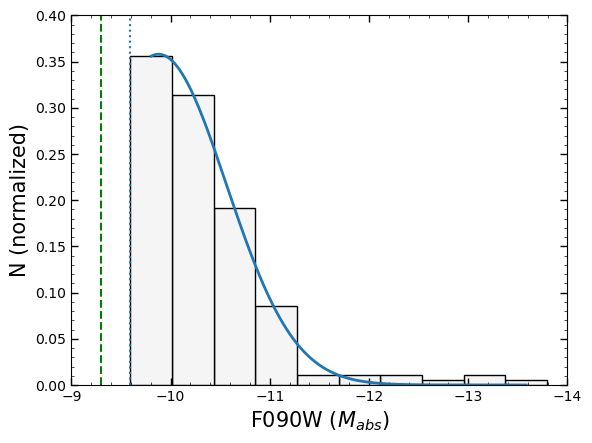}
\caption{Histogram showing the GC candidate LF in the F090W filter. The LF does not reach the turnover point due to the current incompleteness of our rather shallow data. GCs are binned in 0.5-mag bins. The absolute F090W magnitude of the 5$\sigma$ completeness limit is $M_{\rm abs} = -9.59$, as indicated by the vertical blue dotted line. The vertical green dashed line indicates the estimated peak in the F150W GCLF in Abell 2744 by \citet{harris2023jwst}, converted to F090W by using the average color of (F090W--F150W) $\simeq$ +0.47 mag in Fig. \ref{fig:GC_models}a.}
\label{fig:gclf}
\end{figure}

Following other work \citep{harris1991globular, jordan2007acs}, we fit a Gaussian to the GCLF to measure the peak magnitude and standard deviation \citep{harris1991globular, jordan2007acs}. With the mean GC magnitude denoted as $\mu \equiv \langle m \rangle$ and the Gaussian dispersion $\sigma_{m} = \langle \left(m-\mu\right)^{2}\rangle^{1/2}$, we obtain the GCLF fit as
\begin{equation}
    \frac{dN}{dm} = C\exp\Biggl[- \frac{\left(m-\mu\right)^{2}}{2\sigma_{m}^2}\Biggr], 
\end{equation}
where C is an arbitrary amplitude fit as a parameter. Fig. \ref{fig:gclf} shows the GCLF histogram for observed data above the 5$\sigma$ completeness limit as measured in the F090W filter. As shown, we do not quite reach the peak of the GCLF in our data. 
\DELETED{The faintest absolute magnitude we reach in the F090W filter is $M_{\rm abs} = -9.6$ mag. Regardless, our best-fit values are $M_{\rm abs} = -9.87$ and dispersion of $\sigma_m = 0.68$~mag. 
 \citet{harris2014globular} find that their I-band GCLF peaks at M$_I$ $\simeq$ --8.3 mag with a dispersion of $\sigma_m$ $\simeq$ 1.2 mag. This is about 1 mag fainter than what we can currently reach with our shallow VV~191 NIRCam integrations. Using our CMD models, we predict at $z = 0.0513$ an average I--Z color of --0.4 mag, so that the \citet{harris2014globular} LF peak would be at M$_{F090W}$ $\simeq$ --8.7 mag.} More recently, \cite{harris2023jwst} used a best-fitting three-parameter solution to predict the GCLF turnover magnitude in Abell 2744 using data from JWST. They find a predicted turnover magnitude of $M_{F150W} \simeq$ --9.84 mag, but they emphasize that magnitudes fainter than $m_{F150W} \geq$ 30 mag are very uncertain due to the incompleteness of their Abell 2744 GC sample. By using the predicted F150W turnover values found in \citet{harris2023jwst}, we estimate the turnover in our F090W filter to be $-$9.84 + 0.47\,(\textsl{F090W}$-$\textsl{F150W}) $\simeq$ $-$9.4  mag using the average (\textsl{F090W}$-$\textsl{F150W}) color of 0.47 mag (Fig. \ref{fig:f090_f150_CMD_Hist}).
 \deleted{, --9.84 mag + $0.47 (F090W-F150W) \simeq -9.4$ mag.} This estimate is very uncertain due to incompleteness in the VV~191 data. 

Within our current limited GC statistics and S/N-ratios, our best Gaussian fit of the F090W magnitude distribution is $m_{\rm AB}$ $\simeq$ --9.87 mag with a dispersion $\sigma_{m} = 0.68$ mag (see Fig. \ref{fig:gclf}), and given our current S/N limit, with an estimated mean error of $\sim$0.2 mag. Hence, we do not quite reach the GCLF peak at $z = 0.0513$. Additional $\sim$1 mag deeper JWST data would be able to reveal the actual peak in the GCLF at $z = 0.0513$, which has been proposed for future JWST cycles.

\section{Summary and conclusions}

In this study, we have identified a GC sample belonging to the elliptical galaxy VV~191a, and we present JWST photometry using the NIRCam F090W, F150W, F356W, and F444W filters. A brief summary of our findings is as follows:
\begin{enumerate}

\item Using the {\sc Photutils} package DAOStarFinder, we locate starlike point source objects and carefully reject bright background objects that are unassociated with VV~191a. We use photometry to present CMDs to find GC candidates that fall above the 5$\sigma$ completeness limits for analysis. We took great care to account for the sample contamination and reliability. Table 1 and Fig. \ref{fig:v-shape} show our procedure to estimate the number of background galaxies behind VV~191a and how these affect the globular cluster sample in each ringlet in Fig. \ref{fig:starfinder}. The net number of reliable GCs after subtracting background galaxies is 154. After correcting for the cropped portion of the ellipse, the estimated GC counts could be as high as 171.

\item After correcting for the major known sources of (color-dependent) contamination --- the bluest or reddest background field galaxies --- the (F090W--F150W) and (F356W--F444W) color distributions appear to be largely single Gaussian, but with some excess very blue components that comprise 2--4\% of the total objects, and a somewhat larger excess very red components that comprise 5--8\% of the total number of GCs.


\item We use the PARSEC CMD 3.7 GC model to test how well it describes our JWST data. The models show that the VV~191a GCs (above our 5$\sigma$ completeness limit) tend to be up to  $\sim$$10^{6.5} M_{\odot}$. We have chosen a typical range of metallicities that we would expect a GC population to span. We find that the current GC models describe the JWST color range in (F090W--F150W) rather well, but predict (F356W--F444W) colors that are $\sim$+0.3--0.4 mag bluer than the NIRCam data. Future models may need to be updated to improve the modeling of the near-IR colors of old, metal-poor stellar populations. 

\item With PARSEC CMD 3.7 stellar isochrone models, the objects in our analysis are indeed largely GCs in VV~191a and not background galaxies or the brightest stars in the foreground spiral. 

\item Given the current shallow NIRCam detection limits, we find our sample’s GCLF does not reach the expected turnover magnitude. The faintest absolute magnitude we reach in the F090W filter is $M_{\rm abs} = -9.6$ mag. Regardless, our best-fit values are $M_{\rm abs} = -9.87$ and dispersion of $\sigma_m = 0.68$~mag. By using the predicted F150W turnover values found in \citet{harris2023jwst}, we estimate the turnover in our F090W filter to be --9.4 mag, $\pm$ 0.2 mag, using the average (F090W--F150W) color of 0.47 mag.
Deeper photometry would be able to reveal the actual peak in the GCLF.

\end{enumerate}

\begin{acknowledgments}
We dedicate this paper to Susan Selkirk, whose expert graphics skills
helped showcase our VV\,191 images at ASU, NASA, and elsewhere.  
We express our sincere thanks to Alan Dressler and Giovanni Ferrami for their helpful discussions. We would also like to give our thanks to Alex Pigarelli for his time and help with the PSF matching. We thank the referee and also Dr. Giovanni Carraro for very thoughtful suggestions that
helped us improve the submitted manuscript. This work is based on observations made with the NASA/ESA/CSA James Webb Space Telescope. The data were obtained from the Mikulski Archive for Space Telescopes (MAST) at the Space Telescope Science Institute, which is operated by the Association of Universities for Research in Astronomy, Inc., under NASA contract NAS 5-03127 for JWST. These observations are associated with JWST programs 1176 and 2738. R.A.W. acknowledges support from NASA JWST Interdisciplinary Scientist grants NAG5-12460, NNX14AN10G and 80NSSC18K0200 from GSFC. This project was funded by the Agencia Estatal de Investigaci\'on, Unidad de Excelencia Mar\'ia de Maeztu, ref. MDM-2017-0765.
M.N. acknowledges INAF-Mainstreams 1.05.01.86.20. M.A.M. acknowledges the support of a National Research Council of Canada Plaskett Fellowship and the Australian Research Council Centre of Excellence for All Sky Astrophysics in 3 Dimensions (ASTRO 3D) through project number CE17010001. C.N.A.W. acknowledges funding from the JWST/NIRCam contract NASS-0215 to the University of Arizona.
\end{acknowledgments}

{\it Facilities}: Hubble and JWST Mikulski Archive \url{https:// archive.stsci.edu}. Our specific GTO PEARLS observations were retrieved from MAST at STScI, and can be accessed via the following data sets: \dataset[DOI: 10.17909/mdxt-a937]{https://doi.org/10.17909/mdxt-a937}.

{\it Software}: Astropy: \url{http://www.astropy.org} (Astropy Collaboration et al. 2013, 2018); Photutils:
\url{https://photutils.readthedocs.io/en/stable/} (Bradley et al. 2020); PARSEC CMD 3.7: \url{http://stev.oapd.inaf.it/cgi-bin/cmd_3.7}

\bibliography{VV191a_script2.0.bbl}
\appendix

\vspace*{-0.00cm}
\n \section{Adopted VV~191 Distance and its Error}\
\label{secAppA}

We use the close proximity of the spiral galaxy VV 191b to find a total error estimate in its distance. The spiral galaxy VV 191b at $z = 0.0514$ overlaps ---and is in the foreground--- of the elliptical VV 191a at $z = 0.0513$. Neither galaxy shows the kind of asymmetry that suggests significant gravitational interactions within radii $\approx$ 20 kpc, suggesting that they are more than the sum of these radii apart along the line of sight. However, since VV 191a has a lower redshift value than the spiral, the dynamics must be subject to their mutual gravity, with VV 191b falling toward the elliptical (at a velocity of $\sim$30 km/s when taking the redshift difference as a measure of the infall velocity). We, therefore, assume that eventually, the two galaxies may merge, and so will use the difference between the two redshift values ($\Delta$z $\simeq$ 0.0001 or 0.2\%) as a component of the distance error (see section 1, paragraph 7). The major component of the distance error, however, stems from the Hubble tension \citep{collaboration2020planck, riess2022comprehensive}, which amounts to 4.3\% of our adopted midrange value for the Hubble Constant of \Ho $=$ 70 \kmsMpc. Hence, the combined error on the adopted distance to VV~191a/VV~191b is 228 $\pm$ 10 Mpc.

\vspace*{-0.00cm}
\n \section{Method to Assess Color-Dependent Background Galaxy Contaminant Fraction}\
\label{secAppB}

Fig. 10-11 in this section shows in more detail that the color range of BG galaxies is rather similar to that of the likely GC candidates at $z = 0.0513$. As a consequence, NIRCam near-IR color criteria alone cannot straight-forwardly distinguish the BG contaminants from the real GCs candidate. However, we can say how many contaminants are expected in each ringlet sample {\it and} statistically subtracted, but cannot identify {\it which} candidates are the contaminants through their colors alone. When we obtain 8 NIRCam filters for VV~191 in the future, we will attempt to filter out BG galaxy contaminants from their photometric redshifts, but the current four NIRCam filters do not allow us to do this reliably.

Fig. 10a--10c shows how the color distribution of the (F090W--F150W) GC candidates compares to that of the BG field galaxies found within the same color range as described above. In each figure, there is a presence of contaminating background sources. Fig 10a shows a histogram of ONLY the GCs found in the innermost ringlet 1, for which we statistically only have one BG contamination. Since we expect only one galaxy to leak into the GC candidate list in ringlet 1, we can be certain that the small excess of objects at both color extremes in the innermost ringlet is real. We can not, however, identify which single point source is the contaminating one due to the similar colors of the BG galaxies and GCs. Next, Fig. 10b--10c compares the histograms of the GC candidates found within the 3 innermost ringlets and the 5 innermost ringlets to the color distribution of the BG galaxies, respectively. Table 1 shows that there are 8 known BG contaminants expected within the 3 innermost ringlets, and a total of 34 contaminants in ringlets 1--5. A similar analysis holds for Fig. 11a--11c, which shows the color distribution of the GC candidates compared to that of the BG field galaxies in the F356W--F444W filters. In both sets of near-IR filters, the color distribution of BG galaxies is rather similar to that of GC candidates. Hence, we cannot identify with certainty which objects are the contaminants, but we can statistically subtract their expected numbers, as we do in Table 1. In conclusion, Fig. 10--11 shows that the color distribution of background galaxies is very similar to that of our GC candidates and that, especially for ringlets 1--3, the background contamination is modest. That is, the expected background galaxy contamination can thus not easily have caused or explained a major color bimodality, which could have been the case had the field galaxy color distribution been much redder or bluer than that of our GC candidates. As a corollary, the small fraction of outlying objects with very red or very blue colors may be real.

\DELETED{
\begin{figure}[ht]
\centering
\includegraphics[width=0.4\textwidth]{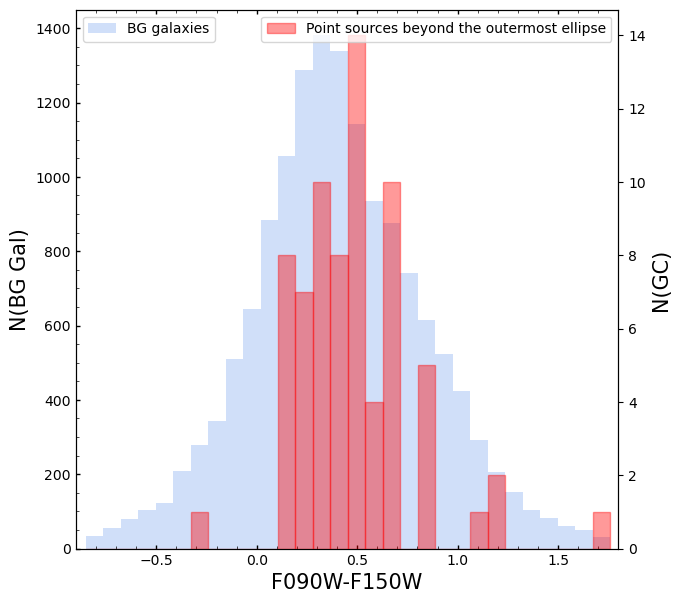}
\includegraphics[width=0.4\textwidth]{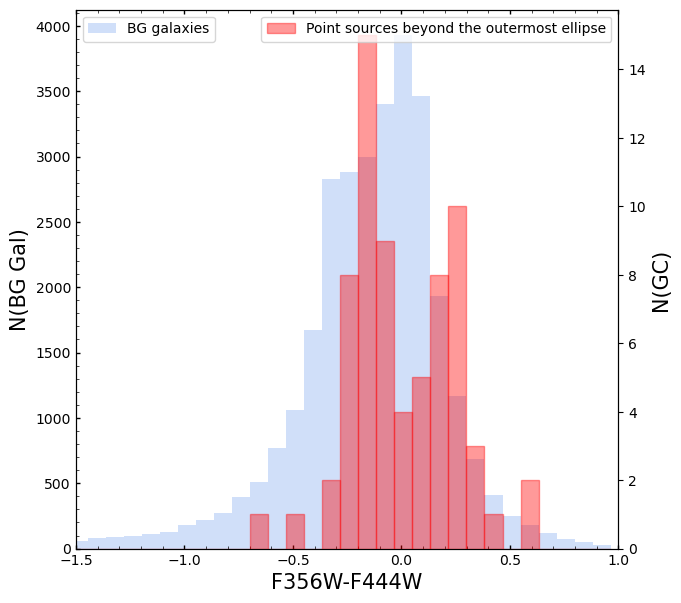}
\caption{Fig. 10a histograms of the BG field galaxies in the (F090W--F150W) filters (blue), which have $m_{\rm AB}$$\leq$ to 27.2 ($\pm$ 0.2) mag as was used for our GC candidate detection. All point-like objects found outside the outermost ringlet are plotted in red. Fig. 10b similarly shows the BG field galaxies in the (F356W--F444W) color blue, which have $m_{\rm AB}$$\leq$ 28.65 ($\pm$ 0.2) mag, where again the point sources found outside the outermost ringlet plotted are plotted in red. We do not use the data from the outermost ringlet 6, and the current plot shows that the colors of BG galaxies are rather similar to those of our GC candidates inside ringlets 1--5.}
\label{fig:outer_hist}
\end{figure}
}

\begin{figure}[ht]
\centering
\includegraphics[width=0.33\textwidth]{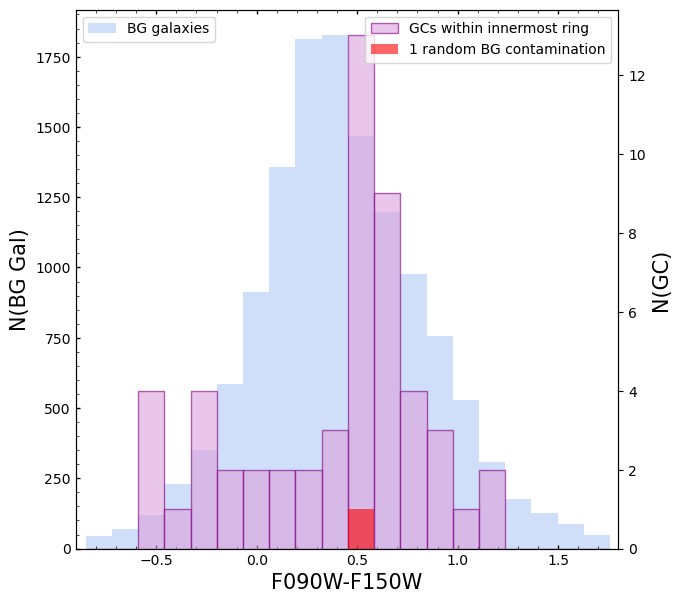}
\includegraphics[width=0.33\textwidth]{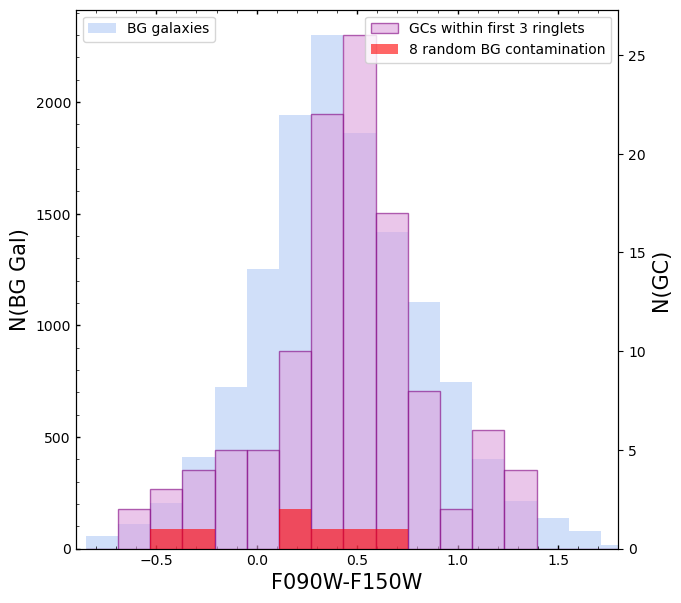}
\includegraphics[width=0.33\textwidth]{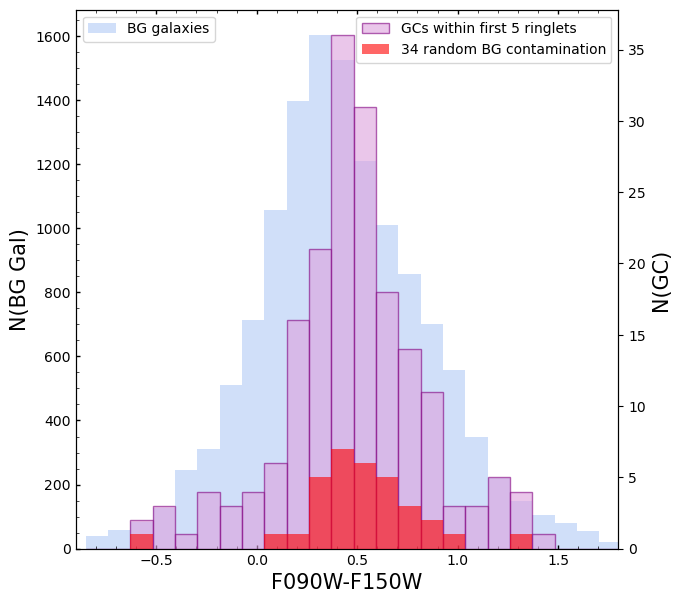}
\caption{Fig. 10a-c shows the BG galaxy color distribution (blue), with GC candidates overplotted (purple) and the BG contamination (red) found in Table 1, Column 4. Fig. 10a shows the color histogram of the GCs found ONLY within the innermost ringlet. Table 1 shows that only $\sim$1 BG contamination has leaked through in the innermost ringlet. We plot a randomly generated BG contamination (red) to visualize its impact. Fig. 10b shows GC candidates found within the 3 innermost ringlets, with 8 randomly generated BG contaminates (\ie\ the sum of rows 1--3 in Table 1, Column 4). Fig. 10c shows the GC candidates in all ringlets that we use, 1--5, with 34 randomly generated BG contaminates (\ie\ the sum of rows 1--5 in Table 1, Column 4).}
\label{}
\end{figure}

\begin{figure}[ht]
\centering
\includegraphics[width=0.33\textwidth]{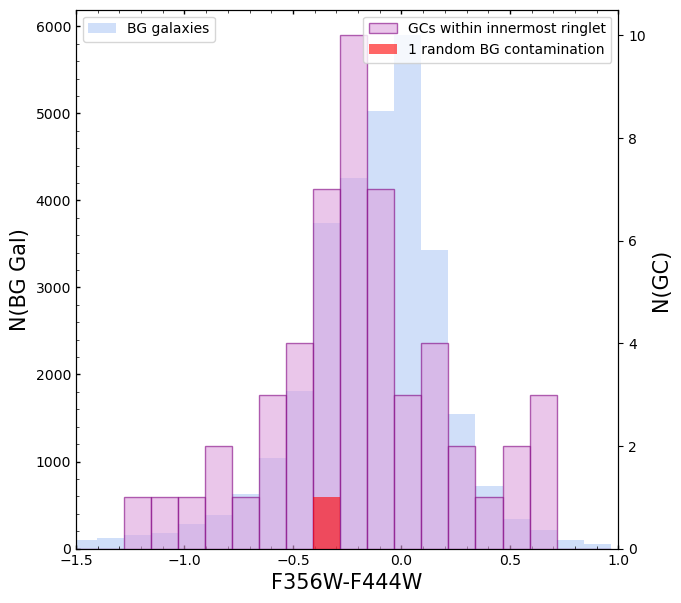}
\includegraphics[width=0.33\textwidth]{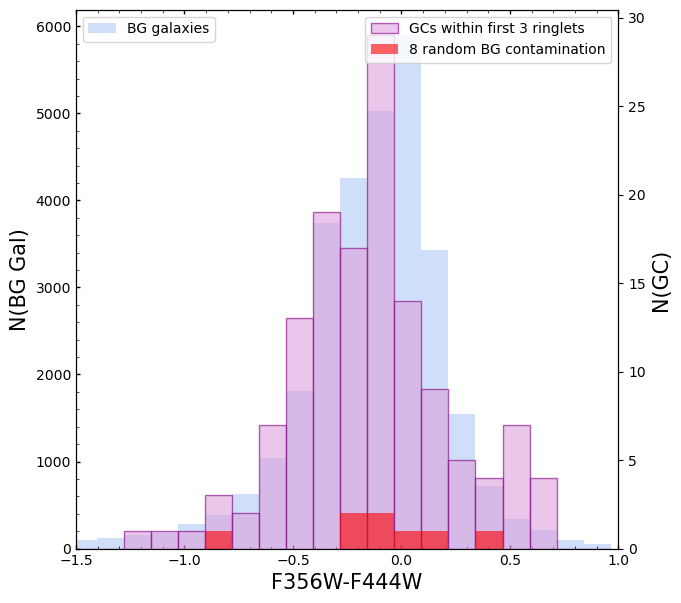}
\includegraphics[width=0.33\textwidth]{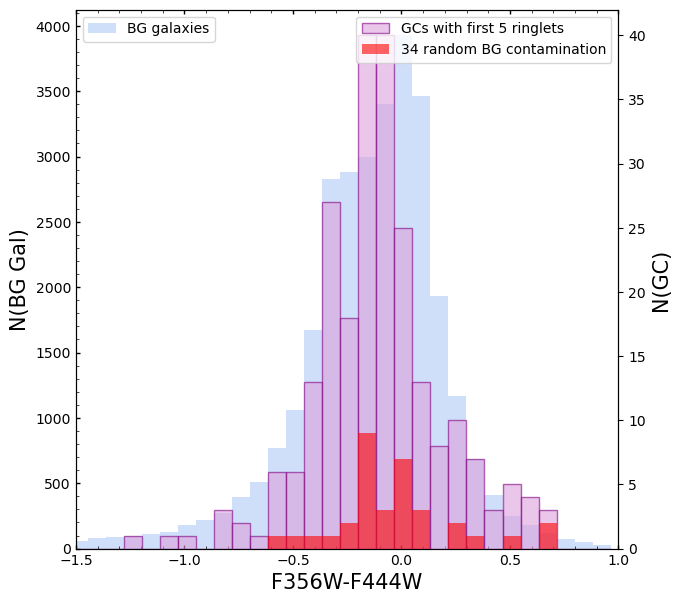}
\caption{Fig. 11a-c shows the BG galaxy color distribution (blue) as explained above in Fig. 10, with GC candidates overplotted (purple) and the BG contamination (red) found in Table 1, Column 4. Fig. 11a shows the color histogram of the GCs found ONLY within the innermost ringlet. Table 1 shows that only $\sim$1 BG contamination has leaked through into the innermost ringlet. We plot a randomly generated BG contamination (green) to visualize its impact. Fig. 11b shows the GC candidates found within the 3 innermost ringlets, with 8 randomly generated BG contaminates (\ie\ the sum of rows 1--3 in Table 1, Column 4). Fig. 11c shows the GC candidates in ringlets 1--5, with 34 randomly generated BG contaminates (\ie\ the sum of rows 1--5 in Table 1, Column 4). Each of these Figures shows that the color range of BG galaxy contaminants is rather similar to that of the  GC candidates. }
\label{}
\end{figure}

\end{document}